\documentclass[twocolumn,english,aps,prb,twocolum,superscriptaddress,bibnotes,amsmath,amssymb,floatfix]{revtex4-1}
\usepackage[colorlinks=true,citecolor=blue,linkcolor=magenta]{hyperref}

\usepackage{changes}
\usepackage{soul}
\usepackage[utf8]{inputenc}
\usepackage[english]{babel}
\usepackage{amsmath,amsfonts,amssymb}
\usepackage[T1]{fontenc}
\usepackage{url}

\usepackage{amsmath}
\usepackage{amsfonts}
\usepackage{amssymb}
\usepackage[version=3]{mhchem}

\usepackage{epstopdf}
\usepackage{graphicx}
\graphicspath{{./Figures/}}

\begin{document}
\title{Nanophotonic soliton-based microwave synthesizers}

\author{Junqiu Liu}
\thanks{These authors contributed equally to this work.}
\affiliation{Institute of Physics, Swiss Federal Institute of Technology Lausanne (EPFL), CH-1015 Lausanne, Switzerland}

\author{Erwan Lucas}
\thanks{These authors contributed equally to this work.}
\affiliation{Institute of Physics, Swiss Federal Institute of Technology Lausanne (EPFL), CH-1015 Lausanne, Switzerland}

\author{Arslan S. Raja}
\thanks{These authors contributed equally to this work.}
\affiliation{Institute of Physics, Swiss Federal Institute of Technology Lausanne (EPFL), CH-1015 Lausanne, Switzerland}

\author{Jijun He}
\thanks{These authors contributed equally to this work.}
\affiliation{Institute of Physics, Swiss Federal Institute of Technology Lausanne (EPFL), CH-1015 Lausanne, Switzerland}
\affiliation{Department of Electrical Engineering, The Hong Kong Polytechnic University, Hong Kong, China}

\author{Johann Riemensberger}
\affiliation{Institute of Physics, Swiss Federal Institute of Technology Lausanne (EPFL), CH-1015 Lausanne, Switzerland}

\author{Rui Ning Wang}
\affiliation{Institute of Physics, Swiss Federal Institute of Technology Lausanne (EPFL), CH-1015 Lausanne, Switzerland}

\author{Maxim Karpov}
\affiliation{Institute of Physics, Swiss Federal Institute of Technology Lausanne (EPFL), CH-1015 Lausanne, Switzerland}

\author{Hairun Guo}
\affiliation{Institute of Physics, Swiss Federal Institute of Technology Lausanne (EPFL), CH-1015 Lausanne, Switzerland}
\affiliation{Current address: Key Laboratory of Specialty Fiber Optics and Optical Access Networks, Shanghai University, 200343 Shanghai, China}

\author{Romain Bouchand}
\affiliation{Institute of Physics, Swiss Federal Institute of Technology Lausanne (EPFL), CH-1015 Lausanne, Switzerland}

\author{Tobias J. Kippenberg}
\email[]{tobias.kippenberg@epfl.ch}
\affiliation{Institute of Physics, Swiss Federal Institute of Technology Lausanne (EPFL), CH-1015 Lausanne, Switzerland}

%\begin{abstract}
%\textbf{.}
%\end{abstract}
\maketitle

\noindent\textbf{\noindent
Microwave photonic technologies \cite{Capmany:07,Marpaung:19}, which up-shift the carrier into the optical domain to facilitate the generation and processing of ultra-wideband electronic signals at vastly reduced fractional bandwidths, have the potential to achieve superior performance compared to conventional electronics for targeted functions.
For microwave photonic applications such as filters \cite{supradeepa2012comb}, coherent radars \cite{Ghelfi:14}, subnoise detection \cite{ataie2015subnoise}, optical communications \cite{temprana2015overcoming} and low-noise microwave generation\cite{Xie:17}, frequency combs are key building blocks. By virtue of soliton microcombs\cite{Kippenberg:18}, frequency combs can now be built using CMOS-compatible photonic integrated circuits\cite{Gaeta:19, Moss:13}, operated with low power and noise, and have
already been employed in system-level demonstrations \cite{Kippenberg:18, Gaeta:19}.
Yet, currently developed photonic integrated microcombs all operate with repetition rates significantly beyond those that conventional electronics can detect and process, compounding their use in microwave photonics.
Here we demonstrate integrated soliton microcombs operating in two widely employed microwave bands, X-band (used e.g. for radar, $\sim$10 GHz) and K-band (used e.g. for 5G, $\sim$20 GHz).
These devices can produce more than 300 comb lines within the 3-dB-bandwidth, and generate microwave signals featuring phase noise levels below $-105$ dBc/Hz ($-140$ dBc/Hz) at 10 kHz (1 MHz) offset frequency, comparable to modern electronic microwave synthesizers.
In addition, the soliton pulse stream can be injection-locked to a microwave signal, enabling actuator-free repetition rate stabilization, tuning and microwave spectral purification \cite{Weng:19}, at power levels compatible with silicon-based lasers (<150 mW) \cite{Liang:10, Morton:18}.
Our results establish photonic integrated soliton microcombs as viable integrated low-noise microwave synthesizers.
Further, the low repetition rates are critical for future dense WDM channel generation schemes\cite{Mazur:19}, and can significantly reduce the system complexity of photonic integrated frequency synthesizers\cite{Spencer:18} and atomic clocks \cite{Newman:18}.}

\def \DWRep {\Delta \omega_{\rm rep}}
\def \DFRep {\Delta f_{\rm rep}}
The synthesis, distribution and processing of radio and microwave signals is ubiquitous in our information society for radars, wireless networks, and satellite communications. With the looming bandwidth bottleneck in telecommunications \cite{hecht2016bandwidth} (due to e.g. future requirements of 5G and the Internet of Things), the tendency is to use carriers in higher frequency bands. As it becomes progressively difficult to generate and digitize electronic signals with increasing carrier frequency, using photonics to process ultra-wideband signals has been extensively explored, commonly referred to ``microwave photonics'' \cite{Capmany:07}. Landmark demonstrations of microwave photonics, such as radar \cite{Ghelfi:14}, analog-to-digital converter \cite{khilo2012photonic}, radio-over-fiber\cite{lim2010fiber}, and waveform generation \cite{khan2010ultrabroad} have achieved bandwidth not attainable using conventional electronics. 
Similarly, the synthesis of low-noise microwave signals, paramount in a large variety of modern applications such as time-frequency metrology \cite{Riehle:17} and wireless broadband communications \cite{rappaport2011state}, have attained unrivalled perfermance in terms of spectral purity (i.e. noise) \cite{Xie:17, Li:14, Liang:15}, by using optical synthesis approaches based on frequency combs \cite{Udem:02, Cundiff:03}.
However, the future deployment of these technologies critically depends on achieving such performance enhancements with photonic \textit{integrated} components\cite{Marpaung:19}.
In this context, chip-scale, integrated-microresonator-based, soliton frequency combs (``soliton microcomb'') \cite{Kippenberg:18} could be key building blocks for microwaves synthesis, and as sources to implement microwave photonic functionalities which require multiple coherent carriers \cite{Torres-Company:14, Wu:18}.
Silicon nitride (Si$_3$N$_4$) \cite{Gaeta:19, Moss:13} - a CMOS-compatible material used as diffusion barrier and etch mask in semiconductor manufacture of integrated circuits  - has given rise to photonic integrated microcombs, which operate with low power and can be integrated with compact lasers \cite{Stern:18, Raja:19} as well as further optical or electrical functionality.
Such integrated Si$_3$N$_4$-based soliton microcombs have been utilized recently in several system-level demonstrations, such as coherent communications \cite{Marin-Palomo:17}, ultrafast ranging \cite{Trocha:18, Suh:18}, astrophysical spectrometer calibration \cite{Obrzud:19, Suh:19}, dual-comb spectroscopy \cite{Dutt:18, Yang:19},  and optical coherence tomography \cite{Marchand:19, Ji:19}, and could form the basis for integrated photonics-based microwave oscillators.
However, the low quality factor and the resulted increasing soliton threshold power with decreasing repetition rates have limited integrated microcombs of repetition rates beyond spectral bands targeted for easy signal processing by regular optoelectronics components (typically $f_{\mathrm{rep}} \gtrsim 100 \mathrm {GHz}$).

\begin{figure*}[t!]
\centering
\includegraphics{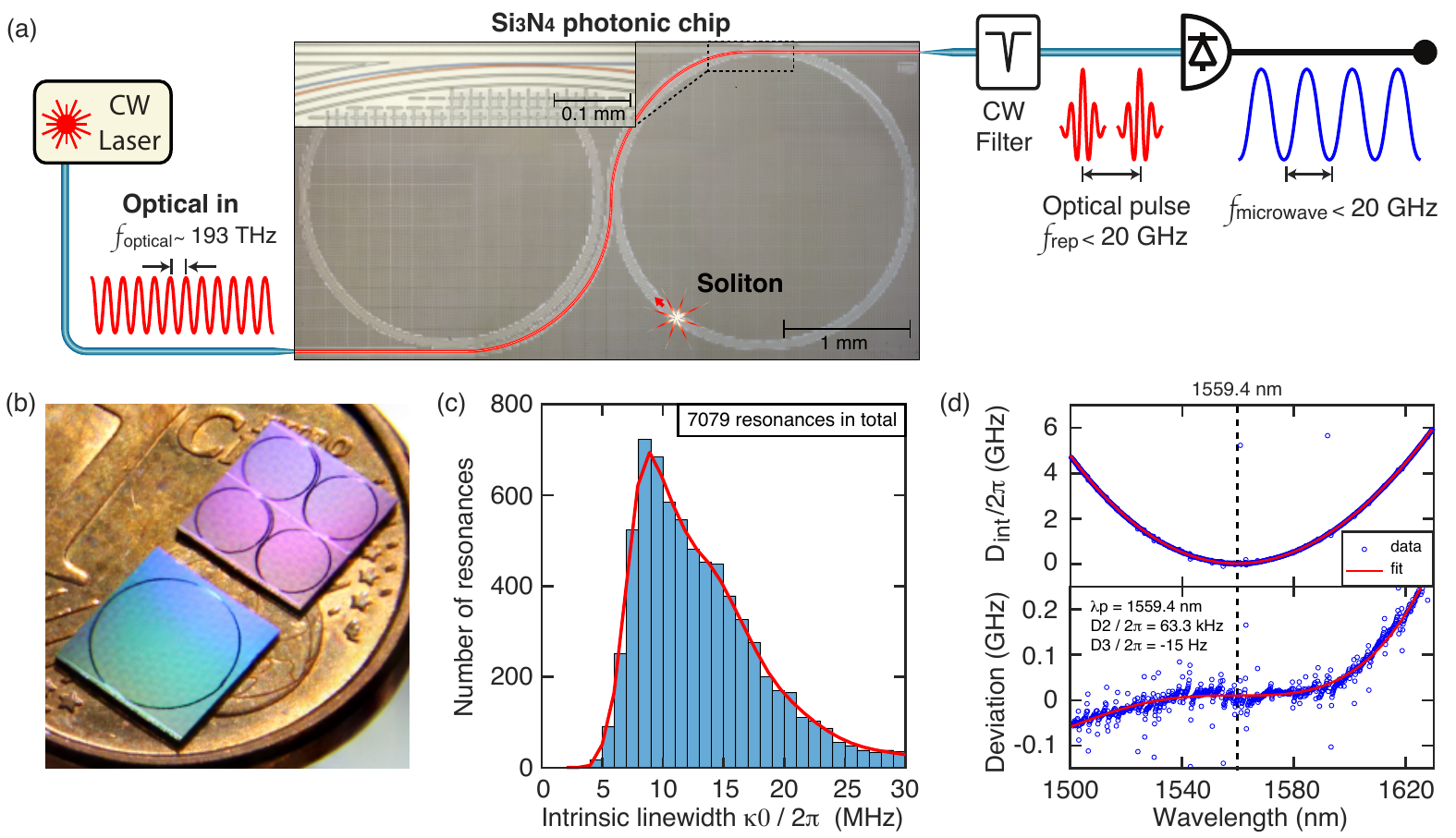}
%\vspace{-0.3cm}
\caption{\textbf{Principle of nanophotonic microwave synthesizers based on integrated soliton microcombs, and characteristics of Si$_3$N$_4$ microresonators}.
(a) Concept of the photonic microwave synthesizer based on an integrated Si$_3$N$_4$ soliton microcomb driven by a CW laser. The microscope image of the Si$_3$N$_4$ photonic chip highlights the bus-waveguide-to-ring-resonator coupling, and stress-release patterns for Si$_3$N$_4$ film crack prevention.
(b) Photograph of Si$_3$N$_4$ photonic chips which are $5\times5$ mm$^2$ in size, in comparison with a 1-cent Euro coin. The chip color is due to the light interference caused by the SiO$_2$ cladding.
(c) Histogram of the intrinsic loss $\kappa_0/2\pi$ of 7079 resonances, fitted with a non-parametric, normal Kernel distribution. The most probable value is $\kappa_0/2\pi=8.5$ MHz, corresponding to $Q_0 > 22\times10^6$.
(e) Top: Fitted microresonator dispersion $D_\text{int}/2\pi$, with $D_1/2\pi\sim19.6$ GHz and $D_2/2\pi\sim63.3$ kHz, referred to $\lambda_\text{p}=1559.4$ nm. Bottom: Resonance frequency deviation from a D$_2$-dominant parabolic profile, defined as $(D_\text{int}-D_2\mu^2/2)/2\pi$, in order to outline mode crossings and $D_3/2\pi\sim-15$ Hz. Mode crossings well below 50 MHz are revealed due the high resonance Q (loaded linewidth $\kappa/2\pi\sim18$ MHz).}
%\vspace{-0.5cm}
\label{Fig:figure1}
\end{figure*}

\begin{figure*}[t!]
\centering
\includegraphics{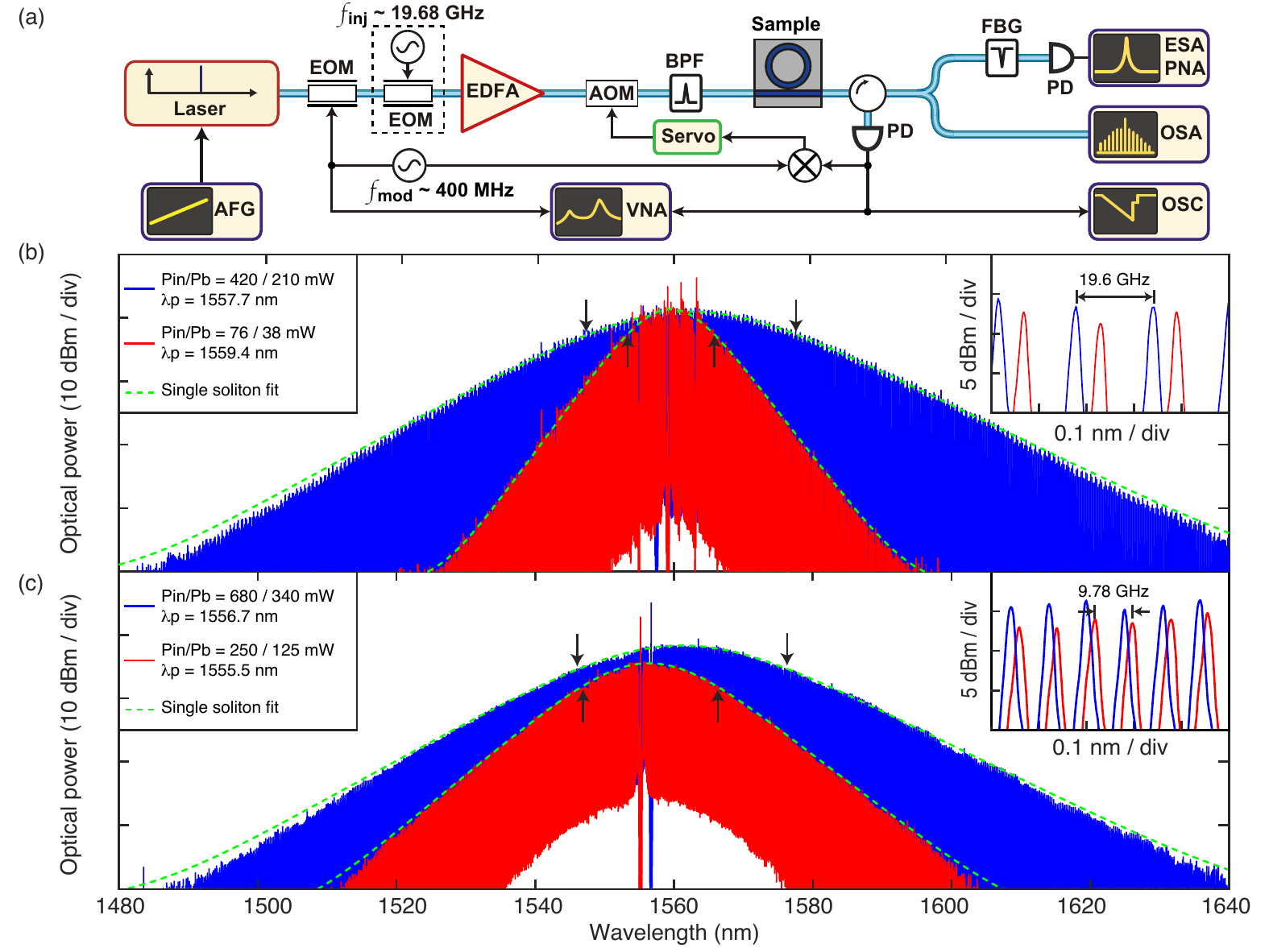}
\caption{\textbf{Single solitons of microwave K- and X-band repetition rates.} 
(a) Experimental setup to generate single solitons and to characterize the soliton phase noise. EOM: electro-optic modulator. AOM: acousto-optic modulator. EDFA: erbium-doped fiber amplifier. AFG: arbitrary function generator. VNA: vector network analyzer. BPF: band-pass filter. FBG: fiber Bragg grating. OSA: optical spectrum analyzer. OSC: oscilloscope. PD: photodiode. ESA / PNA: electrical spectrum analyzer / phase noise analyzer. Dashed box marks the components for soliton injection-locking experiment, which are not used for transmitted power stabilization and cavity-pump detuning stabilization.
(b) Single soliton spectra of 19.6 GHz repetition rate with 38 mW power in sample A (red, 3-dB-bandwidth of 11.0 nm), and with 210 mW power in sample B (blue, 3-dB-bandwidth of 26.9 nm), and their spectrum fit (green). Arrows mark the 3-dB-bandwidths, which contain 69 (red) and 170 (blue) comb lines, respectively. Inset: Spectrum zoom-in showing the 19.6 GHz mode spacing.
(c) Single soliton spectra of 9.78 GHz repetition rate with 125 mW power in sample C (red, 3-dB-bandwidth of 17.4 nm), and with 340 mW power in sample D (blue, 3-dB-bandwidth of 25.8 nm), and their spectrum fit (green). Arrows mark the 3-dB-bandwidths, which contain 139 (red) and 327 (blue) comb lines, respectively. Inset: Spectrum zoom-in showing the 9.78 GHz mode spacing. Note: for soliton spectra in (b) and (c), a BPF is used to filter out the EDFA's amplified spontaneous emission (ASE) noise in the pump laser, and an FBG is used to filter out the pump laser in the soliton spectra.}
%\vspace{-0.5cm}
\label{Fig:figure2}
\end{figure*}

So far, soliton microcombs of repetition rates in the microwave X- and K-band ($f_\text{rep}<20$ GHz) have been demonstrated only on low-index material platforms such as silica and bulk polished crystalline microresonators\cite{Yang:18,Liang:15}, which exhibit limited capability of photonic integration.
On integrated platforms such as Si$_3$N$_4$, the main challenges hindering soliton generation at microwave repetition rates are related to the comparatively low quality ($Q$) factor and thermal effects.
Compared with the widely operated $f_\text{rep}\gtrsim100$ GHz, a laser power up to several Watts is required for microwave repetition rates, not only due to the decreasing finesse, but also the decreasing Q caused by fabrication-related defects such as lithography stitching errors. Stitching errors accumulate with larger pattern areas, thus they can cause serious device failure and low fabrication yield for very long waveguides or large rings. To avoid stitching errors, complex shapes \cite{Johnson:12, Huang:15, Xuan:16} have been utilized for microresonators of 20 GHz free spectral range (FSR), and a high Q exceeding $10\times10^6$ can still be maintained \cite{Xuan:16}. However, the significantly enhanced mode crossings in these microresonators, caused by the spatial mode coupling in the waveguide bending sections, likely prohibit single soliton formation.
Moreover, thermal effects in Si$_3$N$_4$ lead to short soliton steps \cite{Li:17}, thus accessing the single soliton via simple laser frequency tuning is challenging.  
Complex techniques  \cite{Stone:18} can be used to overcome this challenge, at the expense of requiring extra functionalities.
Here, we overcome the above-mentioned challenges, and demonstrate integrated Si$_3$N$_4$ soliton microcombs operating in the microwave X- and K-band,  and use them to build microwave synthesizers which could be utilized for radars and the next generation of wireless networks. 

\textbf{Principle and sample description}: 
The principle of nanophotonic microwave synthesis based on integrated soliton microcombs is depicted in Fig.~\ref{Fig:figure1}(a). A photonic integrated microresonator is driven by a near-infrared continuous-wave (CW) laser to produce an optical pulse stream, which upon photodetection generates a microwave signal, whose frequency depends on the microresonator FSR. 
As the soliton threshold power increases with decreasing FSR, the key challenge here is to generate soliton pulses with power levels compatible with integrated lasers \cite{Liang:10, Morton:18}. We overcome this challenge by using the recently developed photonic Damascene reflow process \cite{Pfeiffer:18a, Pfeiffer:18} to fabricate high-Q integrated microresonators based on ultralow-loss Si$_3$N$_4$ waveguides (linear propagation loss $\alpha\sim1$ dB/m).
Such low waveguide loss is achieved by using several key fabrication techniques, including deep-UV (DUV) stepper lithography based on KrF at 248 nm to pattern waveguides with reduced stitching errors and superior quality (see Methods), as well as stress-release patterns \cite{Pfeiffer:18a} to prevent cracks in the thick Si$_3$N$_4$ film required for strong anormalous GVD. In addition, to minimize the spatial mode coupling between the soliton mode and other waveguide modes, and to optimize Q, we design the microresonators in perfect ring shape, whose diameters are 2.30 (4.60) mm for $\sim$ 20 (10) GHz FSR.
Figure \ref{Fig:figure1}(b) shows a photo of the final Si$_3$N$_4$ photonic chips in $5\times5$ mm$^2$ size, in comparison with a 1-cent Euro coin.
Frequency-comb-assisted diode laser spectroscopy \cite{DelHaye:09} is used to characterize the microresonator dispersion (defined as $D_\text{int}(\mu)=\omega_{\mu}-\omega_0-D_1\mu=D_2\mu^2/2+D_3\mu^3/6+...$, where $\omega_{\mu}/2\pi$ is the frequency of the $\mu$-th resonance relative to the pump resonance $\omega_0/2\pi$, $D_1/2\pi$ corresponds to the FSR, $D_2/2\pi$ is the GVD and $D_3/2\pi$ is the third-order dispersion) and resonance linewidth in the fundamental transverse electric (TE$_{00}$) mode.
For each resonance, the loaded linewidth $\kappa/2\pi=(\kappa_0+\kappa_\text{ex})/2\pi$, intrinsic loss $\kappa_0/2\pi$, and coupling strength $\kappa_\text{ex}/2\pi$ are extracted from each resonance fit. The resonance analysis method is described in Ref. \cite{Liu:16, Liu:18a}. 
Figure \ref{Fig:figure1}(c) shows the histogram of $\kappa_0/2\pi$ of 7079 fitted resonances, from nine characterized 20-GHz-FSR samples. The most probable value is $\kappa_0/2\pi=8.5$ MHz, corresponding to a statistical intrinsic $Q_0>22\times10^6$.
Figure \ref{Fig:figure1}(d) shows the measured $D_\text{int}/2\pi$ and outlines the resonance frequency deviation from a D$_2$-dominant parabolic profile, defined as $(D_\text{int}-D_2\mu^2/2)/2\pi$, in order to reveal mode crossings and the $D_3$ term. As evidenced by Fig. \ref{Fig:figure1}(d), our fabrication and design yield an ideal anomalous GVD with significantly reduced mode crossings compared with the previous works \cite{Johnson:12, Huang:15, Xuan:16}. Details concerning chip input/output coupling and resonator coupling are found in Methods.

\textbf{K- and X-band soliton generation}: 
Using the Damascene reflow process, we fabricate microresonators of FSR in the microwave K- and X- band (sample information in Methods).
Single solitons are generated using simple laser piezo frequency tuning \cite{Guo:16} in all tested samples.
As shown in Fig. \ref{Fig:figure2}(b), in sample A (red), the single soliton is generated with 38 mW power in the bus waveguide on the chip (76 mW power in the input lensed fiber), while parametric oscillation is observed with 7 mW power. 
The single soliton spectrum fit shows a 3-dB-bandwidth of 11.0 nm, corresponding to a pulse duration of 232 fs. Not only is it the first Si$_3$N$_4$ single soliton of a K-band repetition rate, but it also represents an extremely low threshold power for soliton formation, on par with the power values in silica and crystalline microresonators \cite{Liang:15, Yang:18}. This power level is compatible with state-of-art silicon-based lasers \cite{Liang:10, Morton:18}, which makes full integration of on-chip lasers and Si$_3$N$_4$ nonlinear microresonators possible \cite{Stern:18, Raja:19, Volet:18, Suh:19a}, and allows for soliton-based microwave oscillators. 
In sample B, the single soliton is generated with 210 mW power, and features 170 comb lines within the 3-dB-bandwidth of 26.9 nm (94.6 fs pulse duration), ideal for creating dense wavelength-division multiplexing (WDM) channels for coherent communications \cite{Mazur:19}. 
We further generate single solitons of 9.78 GHz repetition rate in the X-band, as shown in Fig. \ref{Fig:figure2}(c), with 125 mW power in sample C (red) and 340 mW power in sample D (blue). The 3-dB-bandwidths are 17.4 nm (red, 139 comb lines, 146 fs pulse duration) and 25.8 nm (blue, 327 comb lines, 98.6 fs pulse duration), respectively. 

\begin{figure*}[t!]
\centering
\includegraphics{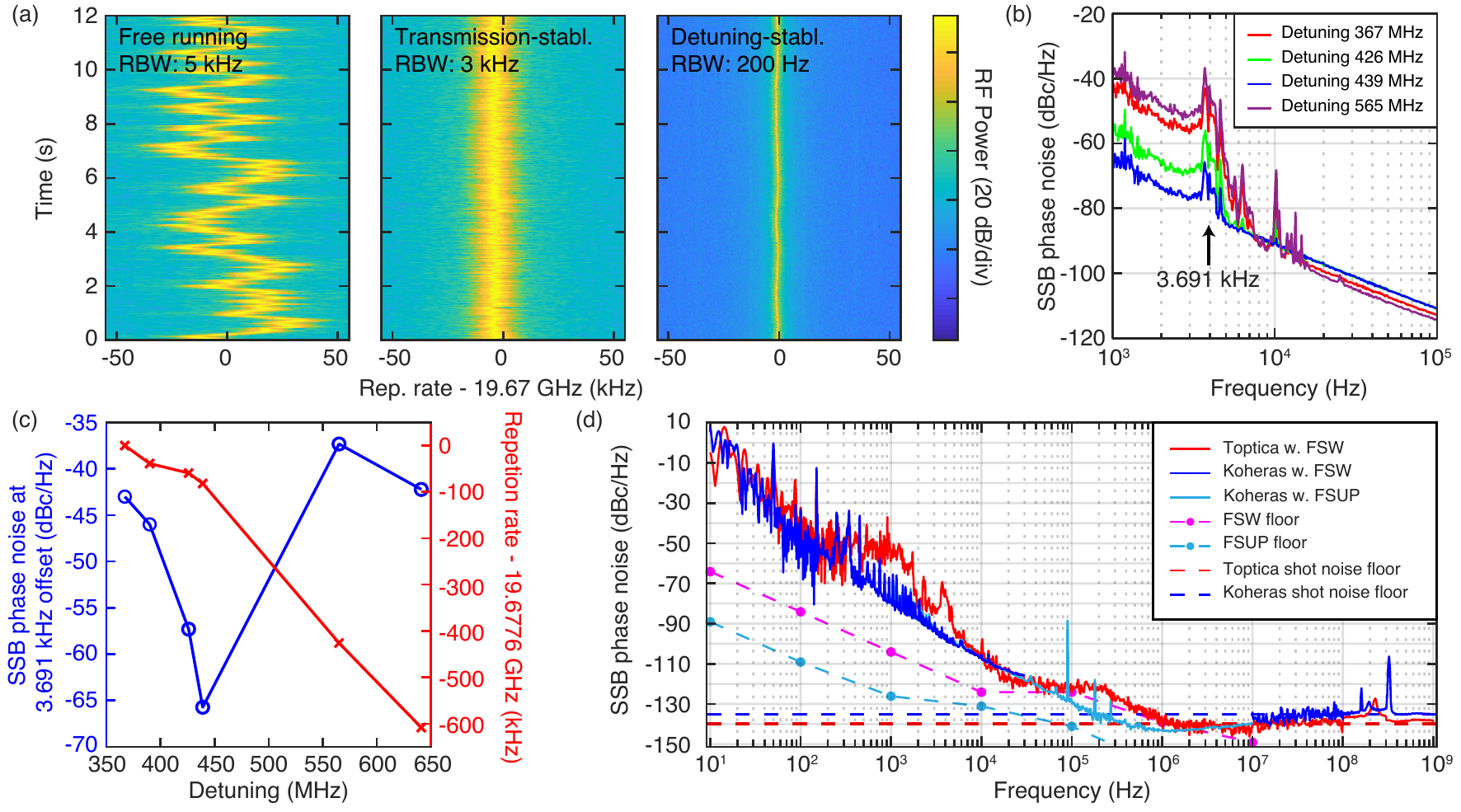}
%\vspace{-0.3cm}
\caption{\textbf{Phase noise characterization of the soliton repetition rate at 19.6 GHz.} (a) Spectrogram of the soliton repetition rate in the free-running state, with transmitted power stabilization, and with cavity-pump detuning stabilization. An oscillation at low frequency $\sim$ 5 Hz is observed in the free running state, and is significantly reduced with transmitted power stabilization or cavity-pump detuning stabilization. (b) SSB phase noise measured at different cavity-pump detunings with power stabilization. A quiet point is observed at the detuning of $\sim$ 439 MHz. (c) SSB phase noise at 3.691 kHz offset Fourier frequency and measured repetition rate shift with different cavity-pump detunings. (d) SSB phase noise measured with the stabilized cavity-pump detuning at $\delta\omega/2\pi\sim$ 400 MHz, using different lasers and PNAs, in comparison with PNA noise floors. The estimated shot noise floors are --140 dBc/Hz for case A with Toptica laser (red), and --135 dBc/Hz for case B with Koheras laser (blue).}
%\vspace{-0.5cm}
\label{Fig:figure3}
\end{figure*}

\begin{figure*}[t]
\centering
\includegraphics{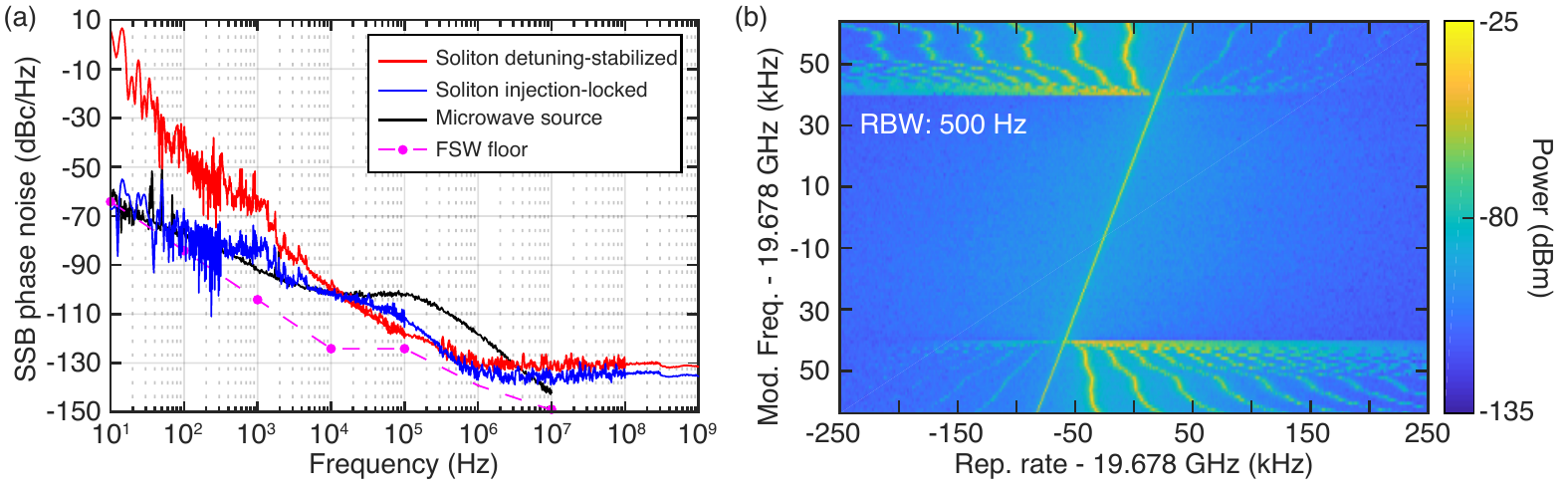}
%\vspace{-0.3cm}
\caption{\textbf{Soliton injection-locking to an external microwave source for improved long-term stability.} (a) SSB phase noise spectrum comparison of the injection-locked soliton, the microwave source used to discipline the soliton, and the soliton with a stabilized cavity-pump detuning. The soliton spectral purification effect is revealed above 10 kHz offset Fourier frequency. (b) Microwave spectrum evolution showing the synchronization of the soliton repetition rate $f_\text{rep}$ with the modulation frequency $f_\text{inj}$ on the CW pump, when $|f_\text{rep}-f_\text{inj}|<$ 40 kHz.}
%\vspace{-0.5cm}
\label{Fig:figure4}
\end{figure*}

\noindent \textbf{Phase noise characterization}:
Next, we perform a thorough analysis of the phase fluctuations of the photonically generated K-band microwave carrier. The measurement setup is shown in Fig \ref{Fig:figure2}(a): the soliton pulse stream is driven by a CW diode laser (Toptica CTL) and the soliton repetition rate is detected on a fast InGaAs photodetector whose output electrical signal is fed to a phase noise analyzer (PNA, Rohde $\&$ Schwarz FSW43). First, the drift of the photodetected soliton repetition rate around 19.67 GHz is characterized in the free-running state. An oscillation at low frequency $\sim$ 5 Hz is observed, as shown in Fig. \ref{Fig:figure3}(a) left, which is likely caused by the unstable chip coupling using suspended lensed fibers, susceptible to vibrations. To mitigate this effect, an acousto-optic modulator (AOM) with a power servo based on a proportional-integral-derivative (PID) controller is used to stabilize the transmitted power through the chip, and compensate for the coupling fluctuations. The previously observed low-frequency oscillation is significantly reduced with stabilized transmitted power, as shown in Fig. \ref{Fig:figure3}(a) middle, which demonstrates that a more robust device coupling scheme is required to improve the soliton stability and the phase noise performance of the soliton-based microwave synthesizer.

The phase noise measurements with different cavity-pump detunings are performed with stabilized transmitted power, as shown in Fig. \ref{Fig:figure3}(b). The detuning is measured using a vector network analyser (VNA) to probe the resonance frequency relative to the laser \cite{Guo:16}. A ``quiet point'' \cite{Yi:17}, caused by mode crossings, is observed at the detuning of $ \delta\omega/2\pi\sim 439$ MHz, and provides the best phase noise performance compared with other detuning values. To evidence the phase noise reduction at the quiet point, the repetition rate shift and the phase noise value at 3.691 kHz Fourier offset frequency, where the laser phase noise exhibits a characteristic feature, are measured with different detunings, as shown in Fig. \ref{Fig:figure3}(c). Note that the quiet point may not be found in every (multi-)soliton state but that, in future works, its presence could be engineered (see Methods). The rest of our measurements are performed out of the quiet point regime.

To further stabilize the soliton-based microwave carrier, we actively stabilize the cavity-pump detuning using an offset sideband Pound-Drever-Hall (PDH) lock \cite {Stone:18} with feedback applied to the pump laser power, which can effectively compensate the cavity resonance jitter induced by coupling fluctuations. As shown in Fig. \ref{Fig:figure3}(a) (right), such detuning-stabilization also stabilizes the soliton repetition rate. Two cases are investigated: In case A, with the power-stabilization, the soliton is driven by a diode laser (Toptica) and the PNA used is the FSW43; In case B, with the detuning-stabilization, the soliton is driven by a fiber laser (Koheras AdjustiK), and, besides the FSW43, an additional PNA (Rohde $\&$ Schwarz FSUP, with cross-correlations) is used only for measuring the 10 kHz - 1 MHz offset frequency range. 
Figure \ref{Fig:figure3}(d) shows the measured phase noise in both cases, as well as the PNA noise floors. In case A, the noise feature within 100~Hz~--~10~kHz offset frequency is caused by the Toptica laser phase noise, while the step-like feature within 20~kHz~--~1~MHz is caused by the FSW43 noise floor, which is the reason why the FSUP is needed to measure this frequency range in case B. Using Koheras with FSW43 and FSUP, case B shows a reduced phase noise, while the phase noise within 200~kHz~--~10~MHz is marginally below the shot noise floor, likely caused by parasitic anti-correlation effects in FSUP~\cite{Nelson:14}. Our analysis shows that, in case B, the main phase noise limitation is the laser relative intensity noise (RIN) for offset frequencies $<1$ MHz, with a contribution from the impact of the thermo-refractive noise (TRN) \cite{Huang:19} in Si$_3$N$_4$ on the detuning within 10 -- 100~kHz offset frequencies  (see Supplementary Information). The absolute single-sideband (SSB) phase noise power spectral density of the microwave carrier shows $\sim$ --80 dBc/Hz at 1~kHz offset Fourier frequency, $\sim$ --110~dBc/Hz at 10 kHz and \mbox{$\sim$ --130~dBc/Hz} at 100~kHz. Note that the phase noise is not measured precisely at a quiet point in case B, therefore further phase noise reduction is possible through quiet point operation (see Methods) and laser RIN reduction.
%A comprehensive comparison of our soliton-based microwave synthesizers with other reported works will be discussed in the conclusion.

\begin{figure}[b!]
\centering
\includegraphics{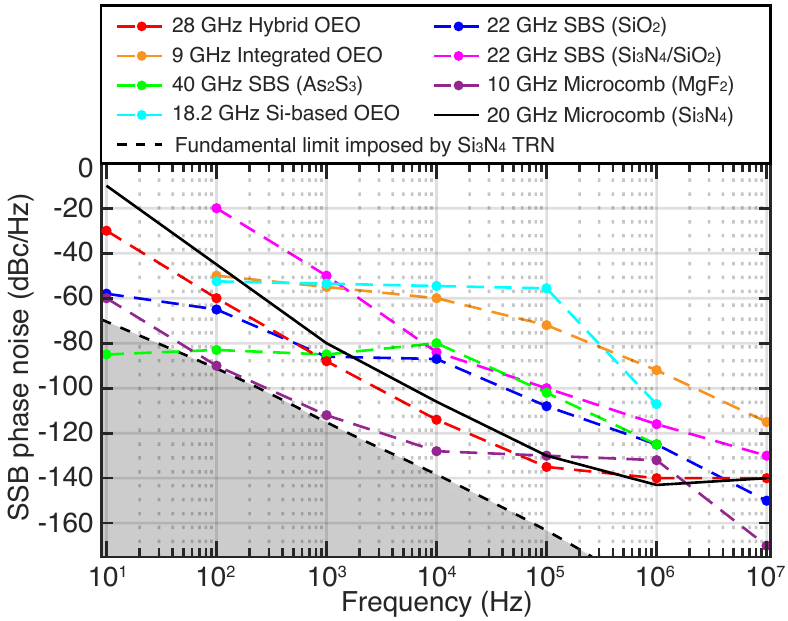}
%\vspace{-0.3cm}
\caption{\textbf{Comparison of demonstrated compact photonics-based microwave oscillators}. 28 GHz hybrid optoelectronic oscillators (OEO) (OEwaves HI-Q$^\text{TM}$ Nano-OEO), 9 GHz integrated OEO \cite{Tang:18}, 40 GHz OEO using stimulated Brillouin scattering (SBS) in As$_2$S$_3$ chalcogenide \cite{Merklein:16}, 18.2 GHz silicon-based OEO \cite{TDo:19}, 22 GHz SBS in SiO$_2$ \cite{LiJ:13}, 22 GHz SBS in Si$_3$N$_4$/SiO$_2$ \cite{Gundavarapu:19}, 10 GHz microcomb in MgF$_2$ \cite{Liang:15}, and 20 GHz microcomb in Si$_3$N$_4$ (this work). The region below the fundamental limit imposed by TRN in Si$_3$N$_4$ \cite{Huang:19} (dashed black) is grey-shaded.}
%\vspace{-0.5cm}
\label{Fig:figure5}
\end{figure}

\noindent \textbf{Soliton injection-locking}:
A variety of microwave photonic applications require long-term stability of microwave signals, represented as low phase noise at low offset frequency.
In our work, the low soliton repetition rate achieved allows soliton injection-locking to an external microwave source \cite{Weng:19}, which can discipline the soliton repetition rate and reduce the low-frequency phase noise. 
A modulation frequency $f_\text{inj}$ swept around 19.678~GHz is applied on the CW pump laser, and the microwave spectrum evolution is shown in Fig. \ref{Fig:figure4}(b). The soliton injection-locking, i.e. synchronization of the soliton repetition rate $f_\text{rep}$ to the modulation frequency $f_\text{inj}$, is observed when $|f_\text{rep}-f_\text{inj}|<40$ kHz. This injection-locking range is more than a 100-fold increase compared to that measured in MgF$_2$ resonators ($\sim300$ Hz in Ref. \cite{Weng:19}), likely caused by the lower Q in Si$_3$N$_4$. The phase noise spectra of the injection-locked soliton (blue), the microwave source used (black), and the soliton with a stabilized cavity-pump detuning (red, as described previously), are compared in Fig.~\ref{Fig:figure4}(a). The phase noise of the injection-locked soliton closely follows the microwave source's phase noise at offset frequency below 10~kHz, apart from a residual bump at 1~kHz which originates from the pump laser. For Fourier offset frequencies above 10 kHz, the soliton-induced spectral purification effect is revealed, as the soliton phase noise departs from the injected microwave phase noise, and becomes similar to the case with only active cavity-pump detuning stabilization. This soliton injection-locking technique can provide extended coherence time for applications such as dual-comb spectroscopy, and allows for coherent combination of microcombs and further scaling of soliton pulse energy.

\noindent \textbf{Conclusion}:
We have demonstrated low-noise nanophotonic microwave synthesizers based on soliton microcombs that operate in the key microwave K- and X-band, with pump power levels compatible with integrated Si-based lasers \cite{Liang:10, Morton:18}.
Figure \ref{Fig:figure5} compares our work to other compact photonics-based microwave oscillators, as well as the fundamental limit of TRN in Si$_3$N$_4$. The measured phase noise in our work is still 30 dB higher than the fundamental limit of Si$_3$N$_4$ TRN (see Supplementary Information), revealing the considerable potential for further phase noise reduction, which can be accessed with improved quiet point operation, device coupling, RIN suppression, and higher Q.
Compared with other microwave oscillators shown in Fig.~\ref{Fig:figure5}, our soliton-based microwave ocillators already show competitive phase noise levels, and represent a critical step towards fully integrated low-noise microwave oscillators for future architectures of radars and information processing networks. Moreover, the low soliton repetition rate achieved here is intrinsically beneficial for future dense WDM channel generation schemes\cite{Mazur:19} and will greatly reduce the complexity of photonic integrated frequency synthesizers \cite{Spencer:18} and atomic clocks \cite{Newman:18}, fostering wide deployment of these technologies in our information society. \\

\noindent\textbf{Methods}
\medskip
\begin{footnotesize}
%\begin{methods}

\noindent \textbf{DUV stepper lithography}: The advantages of DUV stepper lithography over electron beam lithography (EBL), besides the higher yield and lower cost, are: 1. The stitching errors on the wafer are 4 or 5 times smaller than the ones on the reticle used in the industrial standard 4 or 5 demagnification lens systems; 2. The reticle writing using standard photolithography ($\sim$ 1 hour) is much faster than the wafer writing using EBL (> 10 hours), thus the field-to-field (or stripe-to-stripe) time delay is significantly shorter with DUV than with EBL, leading to smaller stitching errors caused by the beam drift; 3. The field (or stripe) size of photolithography for reticle writing is much larger than the field size of EBL for wafer writing, leading to fewer stitching errors; 4. Multipass for reticle writing can be easily adapted with reasonable cost increases. Consequently, DUV stepper lithography can provide superior lithography quality, and has been used in recently demonstrated integrated Brillouin laser \cite{Gundavarapu:19} based on low-confinement Si$_3$N$_4$ waveguides of optical propagation loss below 1 dB/m \cite{Bauters:11, Spencer:14}.

\noindent \textbf{Sample information}: More than 20 samples are tested and single solitons are generated in every sample. Only four selected samples (A, B, C and D) are shown here. Samples A and B are used to generate K-band solitons, as shown in Fig.~\ref{Fig:figure2}(b). Sample A is undercoupled, with a loaded linewidth $\kappa/2\pi\sim18$ MHz and a coupling coefficient $\eta=\kappa_\text{ex}/\kappa\sim1/3$. Sample B is overcoupled, with $\kappa/2\pi\sim27$ MHz and $\eta\sim2/3$. Samples C and D are used to generate X-band single soliton, as shown in Fig.~\ref{Fig:figure2}(c). Sample C is critically coupled, with $\kappa/2\pi\sim17$ MHz and $\eta\sim1/2$. Sample D is overcoupled, with $\kappa/2\pi\sim22$ MHz and $\eta\sim2/3$.
The waveguide cross-sections, width $\times$ height, are $1700\times950$ nm$^2$ for samples A and B, and $2100\times950$ nm$^2$ for sample C and D, respectively. The Raman self-frequency-shift \cite{Karpov:16, Yi:16} is observed of $\sim4.9$ nm in sample B and of $\sim4.4$ nm in sample D. For K-band single soliton generation, further power budget reduction to below 100 mW input requires a higher microresonator Q factor, and may also be achieved by increasing the anomalous GVD ($D_2$) via e.g. coupled resonators exhibiting mode hybridization \cite{Kim:17}, which increases the soliton pulse duration and CW-to-soliton power conversion efficiency.

\noindent \textbf{Coupling scheme}: The microresonator is coupled to a multi-mode bus waveguide of the same cross-section for high coupling ideality \cite{Pfeiffer:17b}. Both the straight and pulley bus waveguides are studied in this work, however no prominent performance difference is observed, likely due to the high Q. Light is coupled into and out of the chip device via double-inverse nanotapers \cite{Liu:18}. The coupling loss is $\sim$ 3 dB per facet, corresponding to $\sim$ 25$\%$ fiber-chip-fiber coupling efficiency. 

\noindent \textbf{Quiet point measurement}: At the quiet point with the detuning $\delta\omega/2\pi\sim439$ MHz, the Raman self-frequency-shift \cite{Karpov:16, Yi:16}, which is the soliton center frequency shift according to $\Omega_{\rm Raman} = - 32 D_1^2 \tau_R\delta\omega^2/15\kappa D_2$ ($\tau_R$ is the Raman shock term), is compensated by the recoil $\Omega_{\rm Recoil}$ caused by a mode-crossing-induced dispersive wave \cite{Yang:16}, such that $ \Omega(\delta\omega) = \Omega_{\rm Raman}(\delta\omega) +  \Omega_{\rm Recoil}(\delta\omega) \approx 0$.
The quiet point is reflected on the repetition rate stability since $2\pi f_{\rm rep} = D_1+D_2\Omega(\delta\omega)/D_1 $.
This point gives the best phase noise performance compared with other detuning values.
However, we emphasize that a quiet point can be engineered via the third-order dispersion $D_3$ that skews the soliton spectrum and leads to a repetition rate shift with detuning \cite{Cherenkov:17}, such that
\begin{align}
2\pi f_{\rm rep} &= D_1 + D_2 \dfrac{\Omega_{\rm Raman}(\delta\omega)}{D_1} \nonumber \\
&\quad + \dfrac{D_3}{3} \left(\dfrac{\Omega_{\rm Raman}(\delta\omega)}{D_1}\right)^2 + \dfrac{1}{3} \dfrac{D_3}{D_2} \delta\omega
\end{align}
Thus the dependence of the repetition rate $f_{\rm rep}$ on the detuning $\delta\omega$ could be mitigated over a broader bandwidth based on a more reliable effect (via $D_3$) than mode crossings. 

\noindent \textbf{Funding Information}: This work was supported by Contract FA9550-19-C-7001 (KECOMO) from the Defense Advanced Research Projects Agency (DARPA), Microsystems Technology Office (MTO), by the Air Force Office of Scientific Research, Air Force Material Command, USAF under Award No. FA9550-15-1-0099, and by Swiss National Science Foundation under grant agreement No. 176563 (BRIDGE).

\noindent \textbf{Acknowledgments}: The authors thank Nils J. Engelsen, Guanhao Huang and Wenle Weng for the fruitful discussion. E.L. and M.K. acknowledge the support from the European Space Technology Centre with ESA Contract No. 4000116145/16/NL/MH/GM and 4000118777/16/NL/GM, respectively. J.H. acknowledges the support provided by Prof. Hwa-Yaw Tam and from the General Research Fund of the Hong Kong Government under project PolyU 152207/15E. H.G. acknowledges the support from the European Union’s Horizon 2020 research and innovation program under Marie Sklodowska-Curie IF grant agreement No. 709249. The Si$_3$N$_4$ microresonator samples were fabricated in the EPFL center of MicroNanoTechnology (CMi).

\noindent \textbf{Author contribution}: J.L. designed and fabricated the Si$_3$N$_4$ samples, with the assistance from R.N.W and H.G.. Samples were characterized and analyzed by J.L. and J.H.. A.S.R., J.L., J.H. and M.K. performed the soliton generation experiment. E.L., A.S.R., J.R. and R.B. performed the phase noise measurement and the soliton injection-locking experiment. J.L., R.B., E.L. and T.J.K. wrote the manuscript, with the input from others. T.J.K. supervised the project.

\noindent \textbf{Data Availability Statement}: The code and data used to produce the plots within this work will be released on the repository \texttt{Zenodo} upon publication of this preprint.
\end{footnotesize}
%\vspace{-0.3cm}
\bibliographystyle{apsrev4-1}
\bibliography{bibliography}
\end{document}

\title{Supplementary Information to: Nanophotonic soliton-based microwave synthesizers}

\author{Junqiu Liu}
\thanks{These authors contributed equally to this work.}
\affiliation{Institute of Physics, Swiss Federal Institute of Technology Lausanne (EPFL), CH-1015 Lausanne, Switzerland}

\author{Erwan Lucas}
\thanks{These authors contributed equally to this work.}
\affiliation{Institute of Physics, Swiss Federal Institute of Technology Lausanne (EPFL), CH-1015 Lausanne, Switzerland}

\author{Arslan S. Raja}
\thanks{These authors contributed equally to this work.}
\affiliation{Institute of Physics, Swiss Federal Institute of Technology Lausanne (EPFL), CH-1015 Lausanne, Switzerland}

\author{Jijun He}
\thanks{These authors contributed equally to this work.}
\affiliation{Institute of Physics, Swiss Federal Institute of Technology Lausanne (EPFL), CH-1015 Lausanne, Switzerland}
\affiliation{Department of Electrical Engineering, The Hong Kong Polytechnic University, Hong Kong, China}

\author{Johann Riemensberger}
\affiliation{Institute of Physics, Swiss Federal Institute of Technology Lausanne (EPFL), CH-1015 Lausanne, Switzerland}

\author{Rui Ning Wang}
\affiliation{Institute of Physics, Swiss Federal Institute of Technology Lausanne (EPFL), CH-1015 Lausanne, Switzerland}

\author{Maxim Karpov}
\affiliation{Institute of Physics, Swiss Federal Institute of Technology Lausanne (EPFL), CH-1015 Lausanne, Switzerland}

\author{Hairun Guo}
\affiliation{Institute of Physics, Swiss Federal Institute of Technology Lausanne (EPFL), CH-1015 Lausanne, Switzerland}
\affiliation{Current address: Key Laboratory of Specialty Fiber Optics and Optical Access Networks, Shanghai University, 200343 Shanghai, China}

\author{Romain Bouchand}
\email[]{romain.bouchand@epfl.ch}
\affiliation{Institute of Physics, Swiss Federal Institute of Technology Lausanne (EPFL), CH-1015 Lausanne, Switzerland}

\author{Tobias J. Kippenberg}
\email[]{tobias.kippenberg@epfl.ch}
\affiliation{Institute of Physics, Swiss Federal Institute of Technology Lausanne (EPFL), CH-1015 Lausanne, Switzerland}

\date{\today}% It is always \today, today,
%\begin{abstract}
%\textbf{.}
%\end{abstract}
\maketitle

\section{Laser phase noise transduction estimation}
\label{sec:PM2PM}
\begin{figure}[b!]
\centering
\includegraphics[width=\columnwidth]{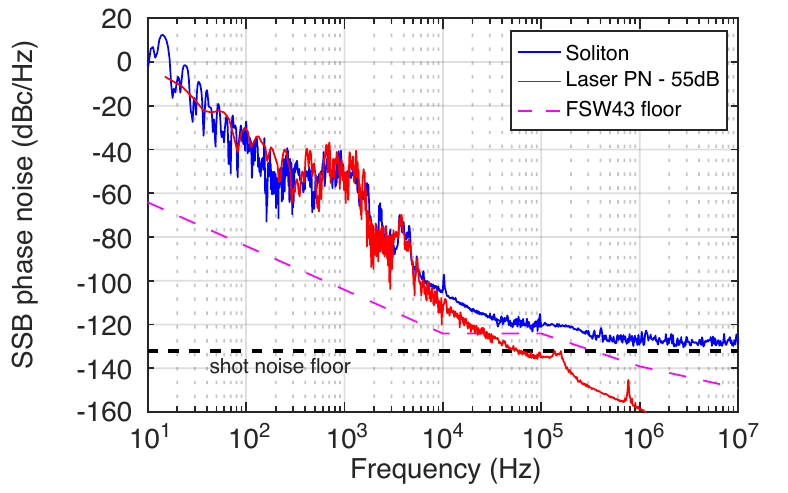}
%\vspace{-0.3cm}
\caption{\textbf{Transduction of the laser phase noise to the microwave phase noise of soliton repetition rate}. The correlation between the optical phase noise of the Toptica laser (red) and the microwave phase noise of the soliton repetition rate (blue) yields an estimation for the PM2PM coefficient of -55 dB.}
%\vspace{-0.5cm}
\label{Fig:SI1}
\end{figure}

The comparison of the laser phase noise (obtained by beating it against an ultra-stable laser) and the phase noise of soliton repetition rate, enables the estimation of the optical to microwave phase transduction (PM2PM). In particular, the Toptica CTL diode laser features a typical noise bump around 3 kHz. Under usual condition, i.e. out of ``quiet point" regime, we measured a PM2PM coefficient of $- 55$ dB, as shown in Fig. \ref{Fig:SI1}.

\section{Laser intensity noise transduction estimation}

\begin{figure}[b!]
\centering
\includegraphics[width=\columnwidth]{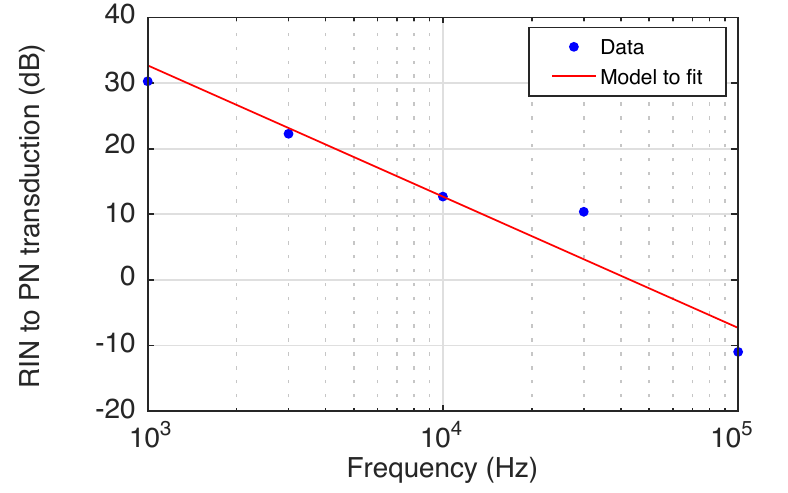}
%\vspace{-0.3cm}
\caption{\textbf{Transduction of the laser RIN into the phase noise of soliton repetition rate}. 
Measured conversion of the pump relative intensity modulation to soliton phase modulation (dots). The red line shows the model $(\alpha/f)^2$ where $\alpha = 79$ Hz/mW.
}
%\vspace{-0.5cm}
\label{Fig:SI2}
\end{figure}

The conversion of laser relative intensity noise (RIN) to the phase noise of soliton repetition rate is experimentally measured. A calibrated pure power-modulation of the pump laser was applied using a 0\textsuperscript{th}-order AOM at frequencies ranging from $10^3$ to $10^{5}$ Hz. The resulting phase modulation strength of the soliton repetition rate is measured by integrating the corresponding peak of the phase noise power spectrum density (PSD). The result of the AM2PM transfer function measurement is shown in Fig. \ref{Fig:SI2}. 

From this measurement, it appears that the AM2PM conversion mostly follows a $1/f^2$ slope (red line in Fig. \ref{Fig:SI2}), meaning that, in fact the amplitude modulation leads to frequency modulation of the repetition rate as:
\begin{equation}
\delta f_{\rm rep} = \alpha \, \delta P_{\rm in}
\end{equation}
We estimate a conversion coefficient of $\alpha = 79$ Hz/mW to match our measurements (the power is defined as the input power in the lensed fiber, before coupled into the chip).
% (the pump fluctuations are considered before the chip facet). 

\section{Sources of oscillator noise}

\begin{figure}[t!]
\centering
\includegraphics[width=\columnwidth]{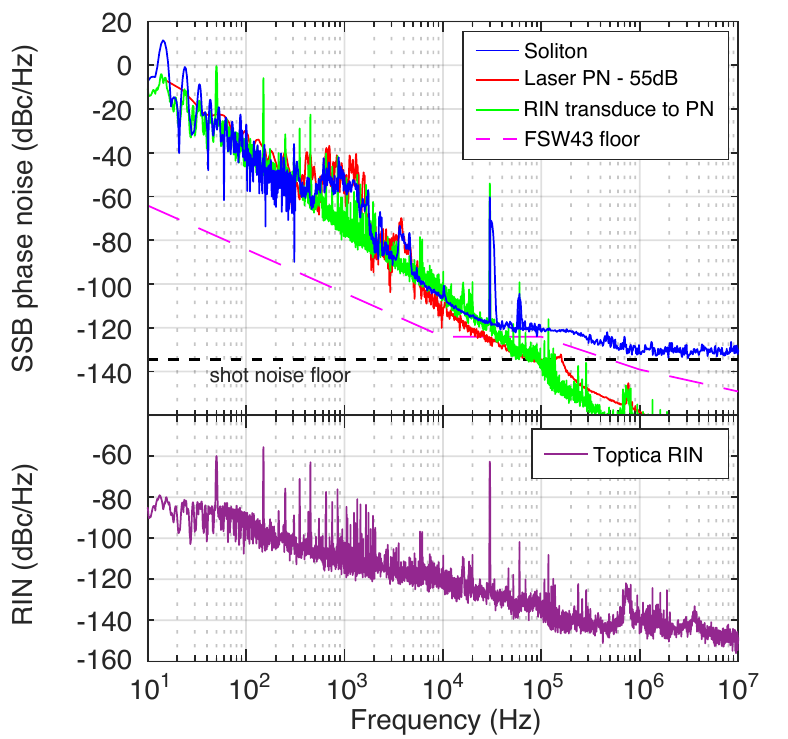}
%\vspace{-0.3cm}
\caption{\textbf{Technical limitations of the soliton phase noise with Toptica diode laser}. The main source of phase noise limitation is the Toptica laser phase noise. The Toptica laser RIN is shown in the bottom panel and its transduction to soliton phase noise is plotted in the top panel (green). The Toptica RIN shows a modulation peak at 10 kHz, which is used to measure the transfer function in Fig. \ref{Fig:SI2}. The black dashed line shows the expected shot noise floor.}
%\vspace{-0.5cm}
\label{Fig:SI3}
\end{figure}

From the previous conversion estimations, we can analyse the origins of the main noise limitations of the soliton repetition rate in our measurements. In the case with Toptica diode laser, as shown in Fig. \ref{Fig:SI3}, the conversion of the laser phase noise and RIN show that the main limiting factor is the laser phase noise, using the previously determined coefficients. Furthermore, FSW43 without cross-correlations is used in this measurement, which limits the phase noise within 30 kHz -- 1 MHz range. At high Fourier frequencies, the shot noise defines the white noise floor.

\begin{figure}[t!!]
\centering
\includegraphics[width=\columnwidth]{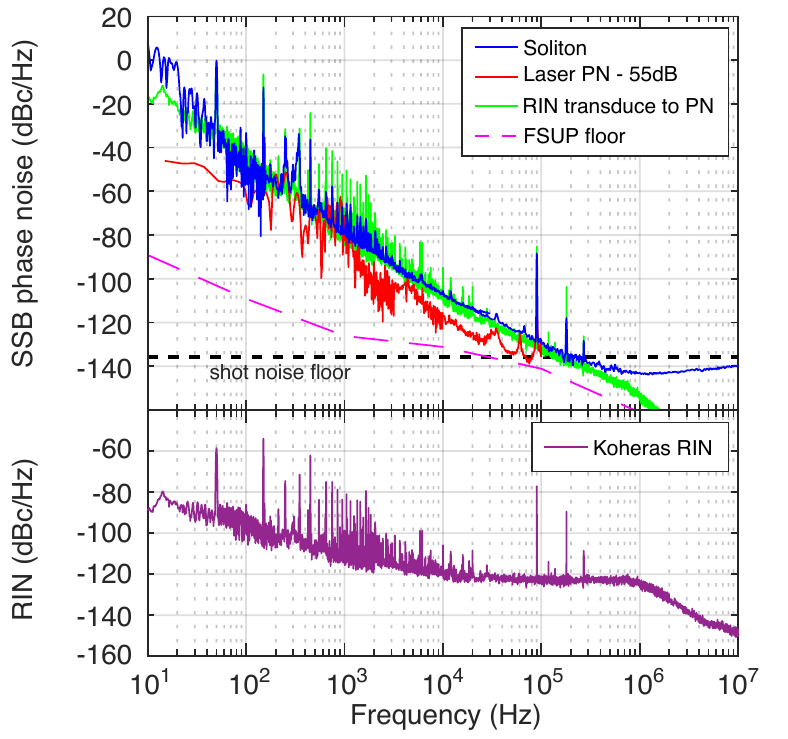}
%\vspace{-0.3cm}
\caption{\textbf{Technical limitations of the soliton phase noise with Koheras fiber laser.} The main source of phase noise limitation is the Koheras laser RIN, as shown in the bottom panel. Its transduction to soliton phase noise is also plotted in the top panel (green), in comparison with the Koheras laser phase noise (red). The black dashed line shows the expected shot noise floor.}
%\vspace{-0.5cm}
\label{Fig:SI4}
\end{figure}

The same analysis with Koheras fiber laser shows that the situation is reversed. As shown in Fig. \ref{Fig:SI4}, the Koheras laser RIN is the main source of phase noise limitation. This is to be expected as fiber lasers typically have lower phase noise than diode lasers. Furthermore, in this measurement, FSUP with cross-correlations is employed to measure the phase noise in the offset frequency range within 30 kHz -- 1 MHz ($10^5$ cross-correlations applied), which alleviates the limitation of the PNA floor. However, this cross-correlation measurement also shows a noise floor that is below the expected shot noise level. We attribute this artefact to a potential correlation between amplitude and phase quadrature of the microwave noise which are known to produce artificially low results~\cite{Nelson:14}. The verification at high offset frequency with FSW43 shows that the signal follows the white shot noise floor.

\section{Resonator thermal noise}

\begin{figure}[b!]
\centering
\includegraphics[width=\columnwidth]{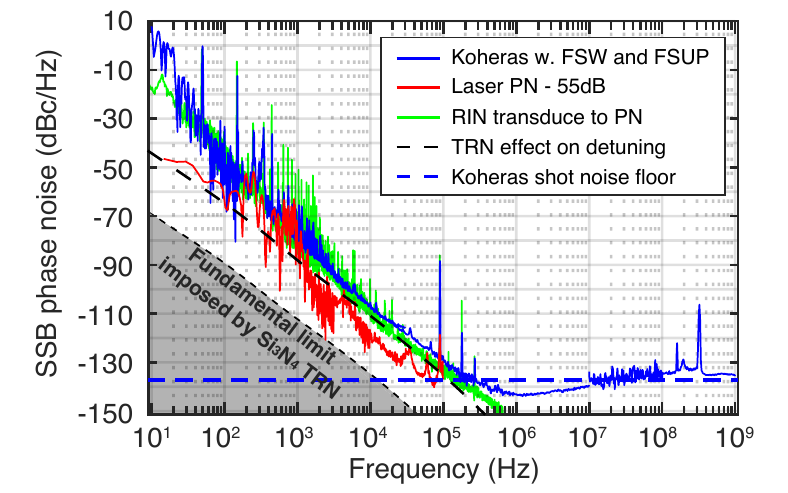}
%\vspace{-0.3cm}
\caption{\textbf{Estimation of the resonator thermal noise impact.} This figure complements to Fig \ref{Fig:SI4}, with the estimated TRN levels on the FSR (fundamental limit) and dutuning (with --55 dB transduction) added. The TRN simulation method is presented in Ref. \cite{Huang:19}.}
%\vspace{-0.5cm}
\label{Fig:SI5}
\end{figure}

The fundamental thermal fluctuations within the optical mode volume of the microresonators lead to the thermo-refractive fluctuations of the resonator FSR~\cite{Kondratiev:18, Huang:19, Drake:19}, and thereby of the cavity resonance (with a magnification by the mode index $m\approx 10^{4}$). While the thermal-induced FSR fluctuations directly affect the soliton repetition rate jitter (which matches the resonator FSR to a first-order approximation), the thermo-refractive noise (TRN) contributes to the phase noise via a second channel. As mentioned in Sec. \ref{sec:PM2PM}, there is a transduction of $\sim$ --55 dB from optical detuning phase noise to the microwave phase noise of soliton repetition rate. Not only the laser contributes to the detuning fluctuations, but also the cavity resonance jitter induced by TRN. This noise can be approximated by multiplying the FSR by the mode number to yield the resonance TRN and substracting 55 dB to account for the PM2PM conversion. Note that the impact of this optical noise can be mitigated by operating at a quiet point (reduction of the PM2PM coefficient), but the FSR fluctuations set a fundamental limit to the phase noise of the microwave oscillator. Numerical simulations to investigate the TRN in Si$_3$N$_4$ microresonators were recently performed and validated experimentally \cite{Huang:19}. Based on the estimated noise for a 20-GHz-FSR microresonator, the estimated impact of TRN is shown in Fig. \ref{Fig:SI5}. The fundamental fluctuation are much lower than the currently measured phase noise level that appear to be limited by the laser RIN at low frequencies and by the optical TRN at higher frequencies.

\bibliography{bibliography}